\documentstyle[twocolumn,aps]{revtex}
\input epsf
\begin{document}
\draft
\wideabs{
\title{Diffusion Dynamics, Moments, and Distribution of First
Passage Time on the Protein-Folding Energy Landscape, with
Applications to Single Molecules}
\author{Chi-Lun Lee,$^1$ Chien-Ting Lin,$^2$ $^*$George Stell,$^2$ and
  $^*$Jin Wang$^{2,3,4}$}
\address{$^1$Department of Physics, State University of New York
  at Stony Brook, Stony Brook, NY 11794 \\ $^2$Department of Chemistry,
  State University of New York at Stony Brook, Stony Brook, NY 11794
  \\ $^3$Global Strategic Analytics Unit, Citibank, One Huntington Quadrangle,
  Suite 1N16, Melville, NY 11747 \\ $^4$Department of Physics, Jilin
  University, Changchun, Jilin 130021, People's Republic of China}

\date{\today}
\maketitle
%\baselineskip= .65cm
%\newpage

\begin{abstract}
We study the dynamics of protein folding via statistical
energy-landscape theory. In particular, we concentrate on the
local-connectivity case with the folding progress described by the
fraction of native conformations. We obtain information for the
first passage-time (FPT) distribution and its moments. The results
show a dynamic transition temperature below which the FPT
distribution develops a power-law tail, a signature of the
intermittency phenomena of the folding dynamics. We also discuss
the possible application of the results to single-molecule
dynamics experiments.
\end{abstract}
 }
  The study of
diffusion along a statistical energy landscape is a very important
issue for many fields. In the field of protein folding, the
crucial question is how the many possible configurations of a
polypeptide chain dynamically converge to a particular folded
state\cite{Levinthal}. Clearly, a statistical description is
needed for a large number of configurational states. According to
the energy-landscape theory of protein
folding\cite{BW,Shak,landscape}, there exists a global bias of the
energy landscape towards the folded state due to natural evolution
selection. Superimposed on this is the fluctuation or roughness of
the energy landscape coming from competing interactions of the
amino acid residues. The folding energy landscape is like a
funnel, and there are in general multiple routes towards the
folded state. Discrete paths emerge only when the underline energy
landscape becomes rough, and local traps (minima) start to appear.
In addition to the thermodynamics of the folding-energy landscape,
the kinetics of folding along the order parameter that represents
the progress of folding towards the native state can be discussed.

When the energy landscape is smooth, the average diffusion time is
a good parameter for the characterization of the dynamical
process. On the other hand, when the energy landscape is rough,
there exist large fluctuations of the energies, and the diffusion
time is expected to fluctuate very much around its mean. In this
case we need to know the full distribution of the diffusion time
in characterizing the folding process.

It is now possible to measure the reaction and folding dynamics of
individual molecules in the laboratory\cite{Moerner,Xie1}. On
complex energy landscapes such as those of biomolecules, reactions
in general do not obey exponential laws, and activation processes
often do not follow the simple Arrhenius relation. However,
measurements on large population of molecules usually cannot
distinguish whether such complex rate dynamics is from the
inhomogeneous distribution among molecules or from intrinsic
features of individual molecules. The study of the statistics of
individual molecular reaction events can greatly clarify these
more subtle reaction processes\cite{intermittency}. The
information on the diffusion-time distribution provides a way to
help unravel the fundamental mechanism of single molecule
reactions. Previously many works have focused on the average rate
behavior\cite{gutin,seno}, whereas very few physical studies and
discussions addressed on the whole distribution of folding rates.
In this paper, we concentrate on the statistics and distributions
of the first passage time (FPT) to probe the folding energy
landscape. A dynamic transition temperature is found above which
the FPT distribution is Poisson-like, and below which the
distribution develops a power-law tail, where non-self-averaging
behavior in kinetics emerges. Moreover, we find that this dynamic
transition temperature is close to the thermodynamic folding
transition temperature.

The framework we adopt here was first introduced by Bryngelson and
Wolynes\cite{BW}. The problem of folding dynamics is modelled as
random walks on a rough energy landscape. In this model, the
energy landscape is generated by the random-energy
model\cite{Derrida}, which assumes that interactions among
non-native states are random variables with given probability
distributions. For this model there are $N$ residues in a
polypeptide chain. For each residue there are $\nu+1$ available
conformational states, one being the native state. A simplified
version of the polypeptide chain energy is expressed as
\begin{equation}
  E = -\sum\epsilon_i(\alpha_i)-\sum J_{i,i+1}(\alpha_i,\alpha_{i+1})
  -\sum K_{i,j}(\alpha_i,\alpha_j)
\end{equation}
where the summation indices $i$ and $j$ are labels of amino acid
residues, and $\alpha_i$ is the state of $i$th residue. The three
terms represent the one-body potential, the two-body interactions
for nearest-neighbor residues in sequence, and the interactions
for residues close in space but not in sequence, respectively. In
this random energy model the energies for the non-native states
and interactions are replaced by random variables with Gaussian
distributions. By this random energy construction one can easily
generate energy surfaces with roughness controlled by adjusting
the widths of these probability distributions. Another feature of
the model is the random-energy approximation, which assumes that
energies for different configurations are uncorrelated. Using a
microcanonical ensemble analysis, the average free energy and
thermodynamic properties of the polypeptide chain can be
obtained\cite{Derrida}.

  In this study, we use the fraction of native conformations $\rho$ as
an order parameter that represents the progress of the folding.
When the kinetics is dominated by the activation folding process,
the states are in general locally connected in $\rho$. Therefore
we assume the local connectivity condition here, assuring that the
dynamics happens continuously with $\rho$. The kinetic process is
carried out with the use of Metropolis dynamics. Therefore the
transition rate from one conformation state to a neighboring state
is determined by the energy difference of these two states, and an
overall constant $R_0$ characterizes the time scale of residue
interactions. The readers are referred to Ref.~[2,11] for detailed
derivations of the dynamic equations. In brief, the energy
landscape is first categorized by the order parameter $\rho$,
along which an energy distribution function is derived via
Eq.~(1). From this energy distribution function, along with the
use of Metropolis dynamics, one can obtain expressions for the
waiting-time distribution function and also the rate distribution
$P(R,\rho)$ for transitions between successive $\rho$'s. Finally,
using continuous-time random walks (CTRW) and assuming the system
to be in quasi-equilibrium, the following generalized
Fokker-Planck equation in the Laplace-transformed space is
obtained\cite{BW}:
\begin{eqnarray}
  \lefteqn{sG(\rho,s)-n_i(\rho) =} \ \ \ \ \ \nonumber \\
  & &\frac{\partial}{\partial\rho}\left\{D(\rho,s)
  \left[G(\rho,s)\frac{\partial}{\partial\rho}U(\rho,s)+
  \frac{\partial}{\partial\rho}G(\rho,s)\right]\right\},
\end{eqnarray}
where $U(\rho,s) \equiv F(\rho)/T + \log[D(\rho,s)/D(\rho,0)]$,
and
\begin{eqnarray}
  F(\rho) &=& N\left\{ -\left(\delta \epsilon - \frac{\Delta\epsilon^2}{2T}
    \right)\rho - \left(\delta L- \frac{\Delta L^2}{2T}\right)\rho^2
    \right. \nonumber \\
   & &\left. + T\rho \log \rho + T(1-\rho)\log \frac{1-\rho}{\nu}\right\}.
\end{eqnarray}
Here $s$, which has a unit of inverse time, is the Laplace
transform variable over time $\tau$, $G(\rho,s)$ is the Laplace
transform of $G(\rho,\tau)$, which is the probability density
function, and $G(\rho,\tau)d\rho$ gives the probability for a
polypeptide chain to stay between $\rho$ and $\rho+d\rho$ at time
$\tau$, while $n_i(\rho)$ is the initial condition for
$G(\rho,\tau)$. $F(\rho)$ is the average free energy for the
polypeptide chain. $T$ is the temperature, $\nu +1$ is the number
of conformational states of each residue, and $\delta \epsilon$
and $\delta L$ are energy differences between the native and
average non-native states for one- and two-body interactions,
respectively. $\Delta \epsilon$ and $\Delta L$ are energy spreads
of one- and two-body non-native interactions. Note that the
two-body energies $\delta L$ and $\Delta L$ already include the
type of interactions due to the second and third term in Eq.~(1).
$D(\rho,s)$ is the frequency-dependent diffusion
coefficient\cite{BW}:
\begin{equation}
  D(\rho,s) \equiv \left(\frac{\lambda(\rho)}{2N^2}\right)\left\langle
  \frac{R}{R+s}\right\rangle_R (\rho) \Bigg/ \left\langle
  \frac{1}{R+s}\right\rangle_R (\rho),
\end{equation}
where $\lambda(\rho) \equiv 1/\nu + (1 - 1/\nu)\rho $. The average
$ \langle \ \rangle_R $ is taken over $P(R,\rho)$, the probability
distribution function of transition rate $R$ from one state with
order parameter $\rho$ to its neighboring states, which may have
order parameters equal to $\rho-\frac 1N$, $\rho$, or $\rho+\frac
1N$. The explicit expression of $P(R,\rho)$ can be found in
Ref.~[2]. The boundary conditions for Eq.~(2) are set as a
reflecting one at $\rho=0$ and an absorbing one at $\rho=\rho_f$.
The choice of an absorbing boundary condition at $ \rho=\rho_f $
facilitates our calculation for the first passage-time
distribution. One can also rewrite Eq.~(2) in its
integral-equation representation by integrating it twice over $
\rho $:
\begin{eqnarray}
  G(\rho,s) &=&
    -\int_{\rho}^{\rho_f}\!d\rho'\int_{0}^{\rho''}\!d\rho''\left[
    sG(\rho'',s)-n_i(\rho'')\right] \nonumber \\
  & &\times \frac{\exp\left[U(\rho',s)-U(\rho,s)
\right]}{D(\rho',s)}
\end{eqnarray}

  The folding time is approximated by the first passage time (FPT)
for the order parameter to reach $ \rho_f $. Thus one can easily
write down the following relation for the FPT distribution
function $P_{FPT}(\tau)$ :
\begin{equation}
  P_{FPT}(\tau) = \frac{d}{d\tau}(1-\Sigma) = - \frac{d\Sigma}{d\tau}
\end{equation}
where $\Sigma(\tau) \equiv \int_{0}^{\rho_f} d\rho\,
G(\rho,\tau)$. The $n$th moment of the FPT distribution function
can be calculated via $\langle \tau^n\rangle = n!
(-1)^{n-1}\int_{0}^{\rho_f}d\rho\, G_{n-1}(\rho)$, where
$G(\rho,s) = G_0(\rho) + s G_1(\rho) + s^2 G_2(\rho) + \cdots$. By
Taylor-expanding Eq.~(5) with respect to $s$, we can solve for
$G_n(\rho)$ and therefore $\langle \tau^n\rangle$ iteratively by
matching the coefficients of $s^n$. In the mean time, one can also
solve for $G(\rho,s)$ directly from Eq.~(5) using the linear
inversion technique. Observing that $P_{FPT}(s) = 1 - s
\Sigma(s)$, where $P_{FPT}(s)$ and $\Sigma(s)$ are Laplace
transforms of $P_{FPT}(\tau)$ and $\Sigma(\tau)$, respectively,
one can calculate $P_{FPT}(s)$ and investigate $P_{FPT}(\tau)$ via
an inverse Laplace-transform technique.

We start the numerical calculations by setting $R_0 = 10^9
s^{-1}$, $N = 100$ and $\nu = 10$, which are approximations to the
physical values \textbf{ We confine ourselves to the single domain
proteins with size less than 100. The proteins with larger size
tend to form domains which is beyond the scope of the current mean
field approximation. } For simplicity we assume $\delta\epsilon =
\delta L$ and $\Delta\epsilon = \Delta L$. Therefore the ratio of
the energy gap between the native state and the average of
non-native states over the spread of non-native states,
$\delta\epsilon/\Delta\epsilon$, becomes an appropriate parameter,
representing the importance of gap bias towards the folded state
relative to the roughness of the landscape. One can show that only
the relative ratios among $\delta\epsilon$, $\Delta\epsilon$ and
$T$ are the controlling parameters in this problem. We set the
initial distribution of the polypeptide chain molecules to be
$n_i(\rho) = \delta(\rho-\rho_i)$, where $\rho_i$ is set to be
0.05. In our calculations we set $\rho_f = 0.9$. This means that
90 percent of the amino acid residues are in their native states.

The mean first passage time (MFPT) $\langle \tau\rangle$ for the
folding process versus a scaled inverse temperature, $T_0/T$, is
plotted in Fig.~1 for various settings of the parameter
$\delta\epsilon/\Delta\epsilon$. We have a inverted bell-like
curve for each fixed $\delta\epsilon/ \Delta\epsilon$, and the
MFPT reaches its minimum at a temperature $T_0$. At high
temperatures, the MFPT is large although the diffusion process
itself is fast (i.e., $D(\rho,s)$ is large). This long-time
folding behavior is due to the instability of the folded state.
The MFPT is also large at low temperature, which indicates that
the polypeptide chain is trapped in low-energy non-native states.
This is in agreement with simulation studies\cite{gutin}.

\begin{figure}[Fig1]
\centering
%\leavevmode
\epsfxsize=2.8in
\begin{center}
%\leavevmode
\epsffile{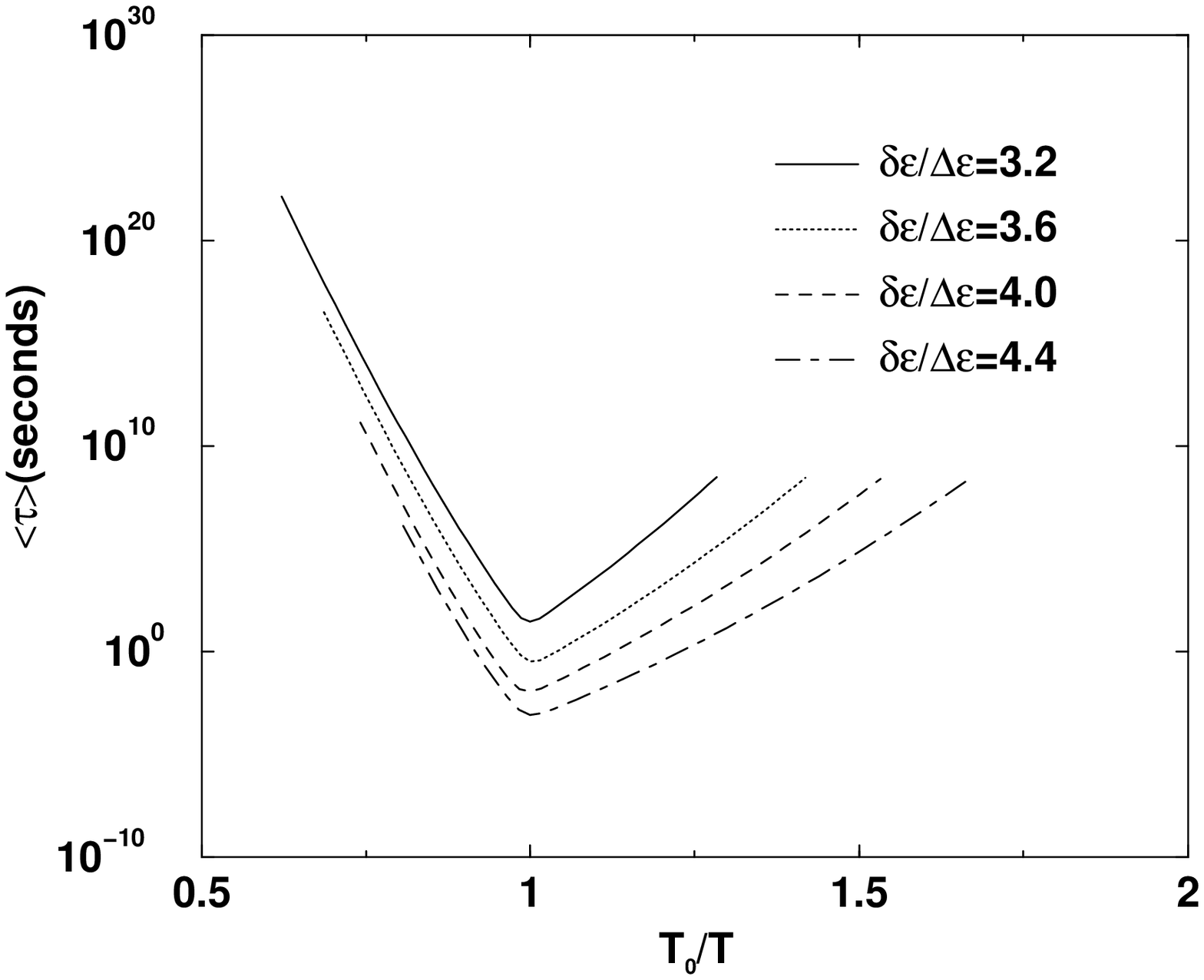}
\end{center}
\caption{MFPT versus reduced inverse temperature $T_0/T$ for
  various $\delta\epsilon/\Delta\epsilon$. As
  $\delta\epsilon/\Delta\epsilon$ increases, the minimum of MFPT decreases.}
\end{figure}

  By comparing the MFPT minimum for various $\delta\epsilon/
\Delta\epsilon$, we conclude that this minimum becomes smaller
when the ratio of the energy gap versus roughness increases. This
suggests that a possible criterion for selecting the subset of the
whole sequence space leading to well-designed fast folding protein
is to maximize $\delta\epsilon/\Delta\epsilon$. In other words,
one has to choose the sequence subspace such that the global bias
overwhelms the roughness of the energy
landscape\cite{Goldstein,Shak,seno}.

  We also calculate the higher-order moments of the FPT distribution.
In Fig.~2 we show the behavior of the reduced second moment,
$\langle \tau^2 \rangle/\langle \tau\rangle^2$. We find that the
reduced second moment starts diverging at a temperature around and
below $T_0$, where the MFPT is at its minimum. This is an
indication of a long tail in the FPT distribution. The divergence
of the second moment also shows that the dynamics exhibits
non-self-averaging behavior.

\begin{figure}[Fig2]
\centering \epsfxsize=2.8in
\begin{center}
\epsffile{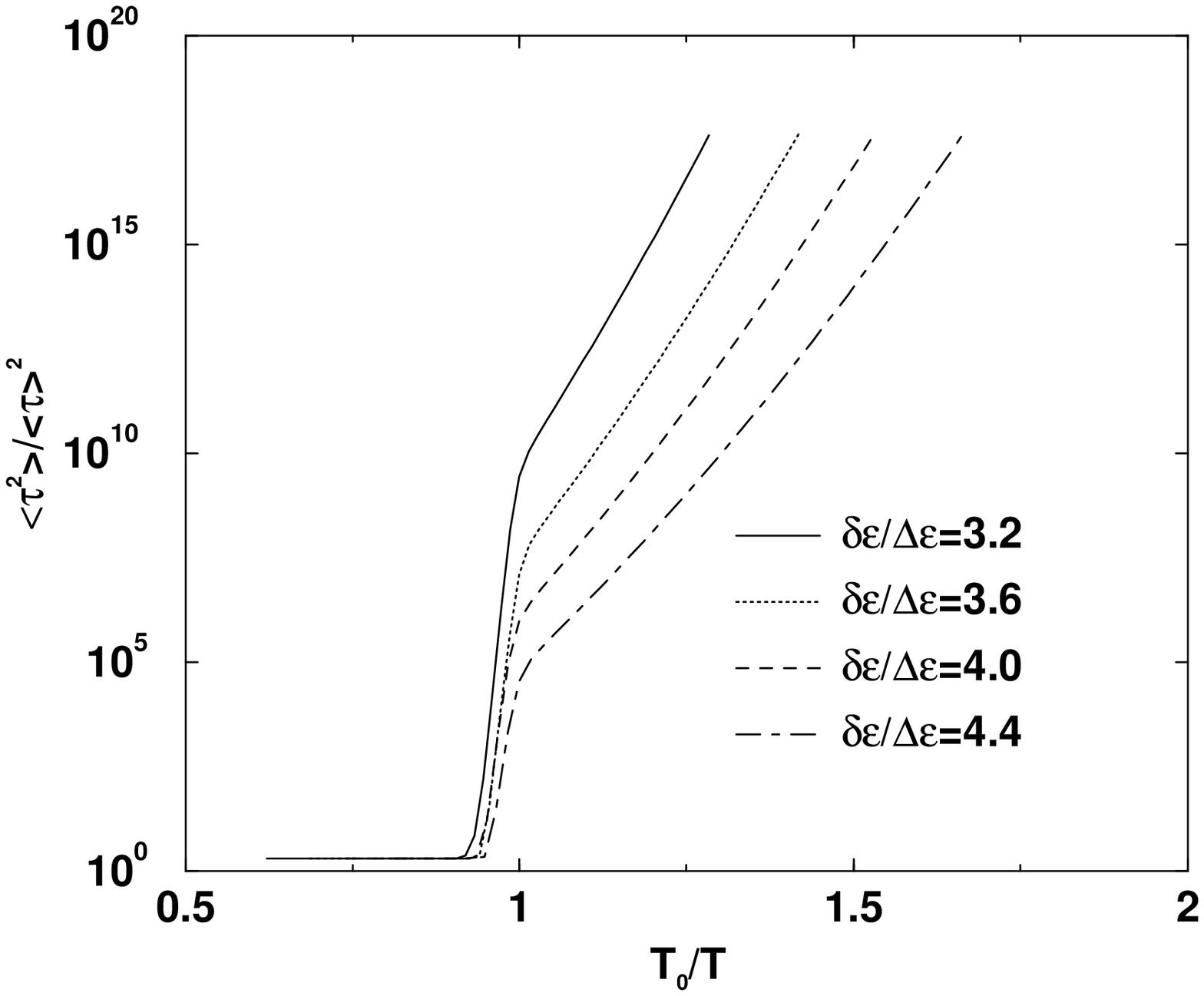}
\end{center}
\caption{$\langle \tau^2\rangle/ \langle \tau \rangle^2$
  versus reduced inverse temperature $T_0/T$ for various
  $\delta\epsilon/\Delta\epsilon$. At high temperature this value keeps
  finite and the folding process is self-averaging. As the temperature drops,
  the value starts to diverge and non-self-averaging behavior emerges. }
\end{figure}

From the study of higher moments, we find the relationship
$\langle \tau^n\rangle \approx n!\langle \tau\rangle^n$ when
$T>T_0$. Therefore in the high-temperature regime the FPT
distribution function is Poissonian and decays exponentially at
large time. When $T < T_0$, it is hard to obtain more information
from the moments because of their diverging behavior. However, we
can study the problem by solving Eq.~(5) directly. By
investigating the behavior of the FPT distribution function in the
Laplace-transformed space, we see that for $T < T_0$ the FPT
distribution is very similar to a L\'evy distribution in time
space, which develops a power-law tail at large time: $
P_{FPT}(\tau) \sim \tau^{-(1+\alpha)}$ for large $\tau$. In Fig.~3
we make a plot of the exponent $\alpha$ versus $T_0/T$ for the
case $\delta\epsilon/\Delta\epsilon = 4.0$. We find that $\alpha$
is decreasing when the temperature is lowered, and $\alpha$
approaches 1 when $T$ goes to $T_0$, where the exponential
kinetics is resumed.

\begin{figure}[Fig3]
\centering \epsfxsize=2.8in
\begin{center}
\epsffile{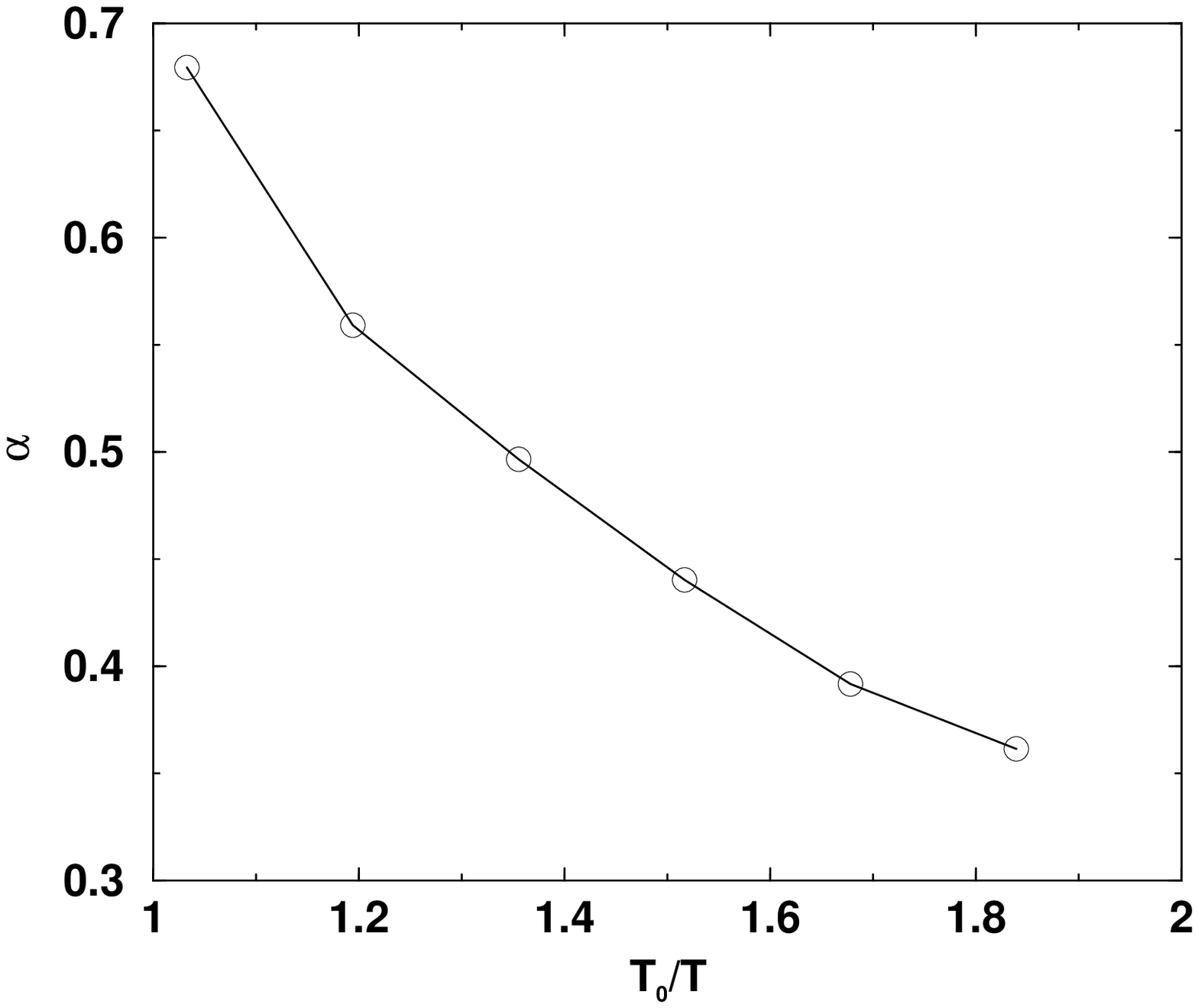}
\end{center}
\caption{The exponent $\alpha$ versus $T_0/T$ for the case
  $\delta\epsilon/\Delta\epsilon = 4.0$ when $T_0/T > 1$. Below the
  transition temperature $T_0$, the FPT distribution is close to a
  L\'evy distribution, which has a power-law tail $P_{FPT}(\tau)
  \sim \tau^{-(1+\alpha)}$ at large $\tau$.}
\end{figure}

From the results above, we find that for a fixed-energy landscape,
there exists a dynamic transition temperature $T_0$. When the
temperature is above $T_0$, the FPT distribution is Poissonian,
indicating exponential kinetics, and in random-walk language we
have normal diffusion on the energy landscape. Below $T_0$, the
variance and higher moments diverge, and the FPT distribution
shows a power-law decay behavior, exhibiting signs of anomalous
diffusion. This indicates the process is non-self-averaging and
distinct folding pathways emerge at various time scales. As a
comparison, we have calculated the thermodynamic
folding-transition temperature $T_f$ by identifying the maximum
heat capacity. We find that $T_f$ is less than but close to $T_0$
for various settings of $\delta\epsilon/\Delta\epsilon$. This
indicates that the thermodynamic and dynamic behavior in proteins
are strongly correlated. \textbf{ However, recent simulation and
experimental results\cite{chan,kuhlman} show that $T_f > T_0$ and
folding is faster below $T_f$ but above $T_0$. This indicates a
possible limitation of the present analytical study. Furthermore,
the crossover behavior near $T_0$ has been shown to be smoother
than what we observe here. This is probably due in part to the
insufficient cooperative interactions in our study \cite{chan}. We
will address these issues in detail in a future publication
\cite{lee}. }

In single-molecule folding experiments, it is now possible to
measure not only the mean but also the fluctuation and moments as
well as the distribution of folding time\cite{singlemole}. Under
different experimental and sequence conditions, one can see
different behavior of the folding time and its distributions. A
well-designed fast folding sequence with suitable experimental
condition exhibits self-averaging and simple rate behavior.
Multiple routes are parallel and lead to folding. A less
well-designed sequence (with larger $\Delta \epsilon$) folds
slowly and often exhibit non-self-averaging non-exponential rate
behavior, indicating the existence of intermediate states or local
traps. In this case, the folding process is sensitive to which
kinetic path it takes, since a slight change in a folding pathway
may cause large fluctuation in the folding time, which indicates
intermittency. One can use single-molecule experiments to unravel
the fundamental mechanisms and intrinsic features of the folding
process. In typical experiments of bulk molecules, it is very hard
to observe and analyze the intermittency, because the dynamics is
averaged over an ensemble of molecules and furthermore, we cannot
say if the bulk phenomena results are either from the intrinsic
features of individual molecules or the inhomogeneous averages
over the molecules.

It is worth mentioning that although we focus on the study of the
protein-folding problem in this paper, the approach we use here is
very general for treating problems with barrier crossings on a
multi-dimensional complex energy landscape. The main ingredient
for this model is Brownian motion on a rough multi-dimensional
landscape, or equivalently, a random walk on a complex network
with a frustration-inducing environment. Since this is quite
general and universal, we expect our results may also be able to
account for a large class of phenomena. In fact the experiments on
glasses, spin glasses, viscous liquids and conformational dynamics
already show the existence of non-exponential distributions at low
temperature. In particular, a recent experiment on single-molecule
enzymatic dynamics\cite{Xie2} shows explicitly the L\'evy-like
distribution of the relaxation time for the underlining complex
protein energy landscape. An interesting study of anomalous
diffusion and non-exponential dynamics has been made recently
using a Fractional Fokker-Planck Equation (FFPE)\cite{barkai} to
describe dynamic processes characterized by L\'evy distributions.
Our results show that the approach we use here can serve as a
microscopic basis for the use of such a FFPE.

The authors thank X. S. Xie for useful discussions. C.-L. Lee and
G. Stell gratefully acknowledge the support of the Division of
Chemical Sciences, Office of Basic Energy Sciences, Office of
Energy Research, U.S. Department of Energy.

\end{document}